\documentstyle[prb,twocolumn,aps]{revtex}
\input{psfig}

\preprint{}
\begin{document}
\draft
\title{Current-voltage relation for superconducting $d$-wave junctions}
\author{Magnus Hurd}
\address{
Department of Microelectronics and Nanoscience, Chalmers University of 
Technology and
G\"oteborg University, S-412 96 G\"oteborg, Sweden\\
and\\
Purdue University, School of Electrical and Computer Engineering,
West Lafayette, Indiana 47907}
\date{January 28, 1997}
\maketitle
\begin{abstract}

We calculate the current-voltage ($I-V$) relation for planar superconducting 
$d$-wave
junctions for both arbitrary transmission of the junction and arbitrary 
orientation of 
the $d$-wave superconductors. 
The midgap
states (MGS) present at interfaces/surfaces of a $d$-wave
superconductor influence the $I-V$ relation.
In some arrangements we find considerable negative differential conductance and
lower threshold voltage for non-zero current due to resonant conduction
through MGS.
\end{abstract}

 
\narrowtext

Soon after the discovery of the high-$T_c$ cuprates it was realized that 
these materials did not
in a straightforward way conform to the conventional BCS scheme.
Therefore, a number of researchers started to speculate whether the
symmetry 
of the order parameter in the high-$T_c$ cuprates might deviate from the
spherical symmetric one ($s$-wave) found in the conventional superconductors.
Recently, an already lively discussion on the subject has become even more 
intensified.
Quite a few experiments probing the magnitude of the order 
parameter\cite{Harl} do
at least not exclude the possibility of an
order parameter with symmetry different from $s$-wave. In fact, some
experiments sensitive to the phase of the order parameter are claimed to 
be impossible
to understand without the assumption of $d$-wave symmetry of the 
order parameter.\cite{Woll,Tsue} On the other hand, there are papers claiming 
that
there is a measurable (if not dominating) $s$-wave component of the order 
parameter.\cite{ChauLin,Sun}
Besides the experimental activity, there are 
several microscopic theories formulated that advocate 
$d$-wave symmetry, in part using arguments based on the strong magnetic 
interactions present in the cuprates.\cite{Harl}

In addition to the discussion above, Hu has pointed out a possible connection
between the often observed zero-bias conductance peaks in high-$T_c$ junctions
and the zero-energy bound states or midgap states (MGS) theoretically 
predicted to be
present at interfaces/surfaces of $d$-wave superconductors.\cite{Hu}
The origin of MGS is due to normal scattering at the interface/surface. A
quasiparticle in the superconductor changes its momentum when scattered. 
Therefore, the quasiparticle in general experiences a different order
parameter before and after the scattering event, since in the $d$-wave case the
order parameter depends on the momentum. Due to Andreev reflection the
quasiparticle will retrace its path and form a closed loop, leading to a bound
state when the order parameter before and after scattering differs in 
sign.\cite{Hu}
\begin{figure}
\centerline{\psfig{figure=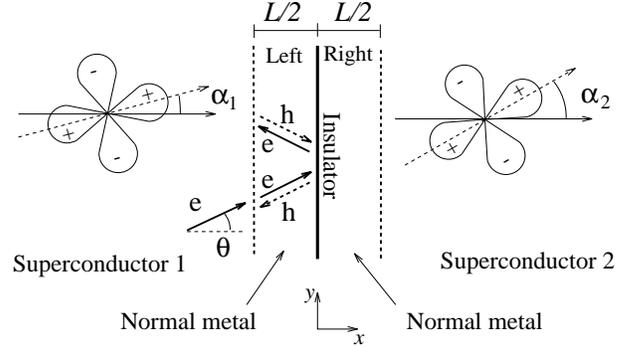,width=8cm}}
\vspace*{3mm}
\caption{
Layout of the junction in the $ab$-plane. The $d$-wave superconductors 
"1" and "2" can be 
rotated an angle $\alpha_{1,2}$. 
A quasiparticle is incident on the barrier
at an angle $\theta$. The length $L$ is only conceptual and set to zero in this
paper.
}
\label{fig1}
\end{figure}
Calculations of $I-V$ curves for normal metal-superconductor (NS) junctions 
indeed confirmed the importance of MGS in the $d$-wave
case.\cite{TanKas,XMT} Due to conduction through MGS, a peak at zero voltage 
grows up in the conductance as 
the transmission of the junction is decreased, which is different from the 
$s$-wave case. 

Motivated by both the general discussion regarding the underlying symmetry of
the order parameter of the cuprates and the new understanding of MGS in $d$-wave
junctions we calculate in this paper the current-voltage ($I-V$) relation for
junctions with anisotropic superconductors. 
To
produce numerical results allowing us to compare our calculation to 
earlier calculations 
we specialize to the case of $d$-wave symmetry. Therefore 
we choose the order parameter to be 
$\Delta(\theta)=\Delta_0\cos(2(\theta-\alpha))$,
where $\theta$ is the angle of incidence for a quasiparticle approaching the
junction and $\alpha$ is the angle of orientation of the $d$-wave
superconductor with respect to the interface as explained in Fig.~\ref{fig1}. 
The cases $\alpha=0$ and $\alpha=\pi/4$ correspond to $d_{x^2-y^2}$ and
$d_{xy}$ symmetry, respectively.
For the $s$-wave case, $\Delta(\theta)=\Delta_s$.

There are some calculations of $I-V$ curves for $d$-wave 
junctions in the literature.\cite{WenOst,DevFul,BOZ,BGZ} Using the tunneling 
hamiltonian
formalism,\cite{WenOst,BOZ} one can calculate the conductance as a 
convolution of the 
density of states in the $d$-wave superconductors. Unmodified, such an 
approach cannot describe the action of MGS. Now, in some configurations of
$d$-wave superconductors there are no MGS, and we can therefore compare the
curves with our calculations (in the tunneling limit) as we will do later 
in this paper. Refs.\ \onlinecite{DevFul} and \onlinecite{BGZ} are using 
methods somewhat similar to ours; however, the presence of MGS is not
discussed. There is also work on asymmetric (different gaps of the
electrodes) SNS junctions in point-contact geometry (one or few conducting
modes in the normal region).\cite{HDB1,HDB2} This theory applies straight to
the $d$-wave case since for point-contacts the modes making it through the
contact correspond to $\theta\approx 0$. In this case the MGS are absent.
However, for the planar junction quasiparticles with any angle $\theta$ of 
incidence come into play, making it necessary to consider the MGS. 

When there are superconductors on both sides of the junction multiple 
Andreev reflection (MAR) will take place.\cite{BSW,AveBar}
In this paper we use methods recently developed to calculate $I-V$ curves of
SNS junctions for arbitrary
transmission in the $s$-wave case.\cite{BSW,AveBar}
These methods will
automatically take into account any effects from the MGS, if appropriately
modified. The implementation of these modifications is 
our new contribution, as presented in this paper.     

Our
calculation is two-dimensional, using a cylindrical Fermi surface.\cite{TanKas} 
Furthermore, the normal scattering will
change the angle (or momentum) of an quasiparticle from 
$\theta$ to $\bar{\theta}=\pi-\theta$. 
This simply
means that the quasiparticle experiences a different pair potential before
and after the scattering event, since in
general $\Delta(\theta)\neq\Delta(\bar{\theta})$. (The exception 
is the case 
$\alpha=0$.) Therefore, one has to book-keep the two different 
gaps in the calculation.\cite{Hu,TanKas,XMT} 

The system we study is shown in Fig.~\ref{fig1}.
The regions
"Left" and "Right", separated by a barrier with transmission amplitude
$t(\theta)$ and reflection amplitude $r(\theta)$, are only 
conceptual since we will only study the case when the
normal region is short (zero-length limit). We model the barrier using a
$\delta$-function potential to derive the angular dependence of $r$ and
$t$.\cite{Brud} 

The time-dependent Bogoliubov-de Gennes equation for 
anisotropic superconductors is 
\begin{eqnarray}
&&\int d{\bf y}
\left\lgroup
\begin{array}{cc}
h_0({\bf x},{\bf y}) & \Delta({\bf x},{\bf y},t) \\
\Delta^*({\bf x},{\bf y},t) & -h_0({\bf x},{\bf y})
\end{array}
\right\rgroup
\left\lgroup
\begin{array}{c}
u({\bf y},t) \\
v({\bf y},t)
\end{array}
\right\rgroup
= \nonumber \\
&&i\hbar\frac{\partial}{\partial t}
\left\lgroup
\begin{array}{c}
u({\bf x},t) \\
v({\bf x},t)
\end{array}
\right\rgroup,
\; \ \ h_0({\bf x},{\bf y})=\delta({\bf x}-{\bf y})(\frac{p_{y}^{2}}{2m}-\mu).
\label{tdBdG}
\end{eqnarray}
In the equation above, $\mu$ is the
chemical potential, which is $\mu_1$ ($\mu_2$) in superconductor 1 (2). Since 
there is a voltage $eV$ between the two 
superconductors, the chemical potentials are different , $eV=\mu_2-\mu_1$. Due
to this difference, the phase $\phi$ of the order parameter 
$|\Delta|e^{i\phi}$ depends on time according to the Josephson relation 
$\partial_t\phi=2eV/\hbar$, leading to inelastic scattering. (For details, 
see Ref.\ \onlinecite{HDB1}.)

Solving Eq.\ (\ref{tdBdG}) piecewise in each region, we follow the standard 
treatment previously 
used in a number of papers.\cite{Hu,TanKas,XMT} To this end, the center of mass
coordinate ${\bf R}=({\bf x}+{\bf y})/2$ and the relative coordinate 
${\bf r}={\bf x}-{\bf y}$ are introduced, expressing the order parameter as 
$\Delta({\bf r},{\bf R},t)$. 

We will in the following assume approximate solutions to Eq.\ (\ref{tdBdG}) 
of the form (in each region)
\begin{eqnarray} 
\left\lgroup
\begin{array}{c}
u({\bf x},t) \\
v({\bf x},t)
\end{array}
\right\rgroup
=
\sum_{\bf k}
e^{i{\bf k}\cdot {\bf x}}
\left\lgroup
\begin{array}{c}
u_{\bf k} \\
v_{\bf k}
\end{array}
\right\rgroup
e^{-iE_{\bf k}t}
\; .
\end{eqnarray} 
We approximate the integral operator of 
Eq.\ (\ref{tdBdG}) by neglecting terms of the order 
$(k_F\xi_0)^{-1}$ (the quasiclassical approximation):\cite{Brud}
\begin{equation}
\int d{\bf y} \Delta({\bf x},{\bf y})v({\bf y},t)\approx
\sum_{\bf k} \Delta({\bf k},{\bf x})v_{\bf k}e^{i{\bf k}\cdot
{\bf x}}e^{-iE_{\bf k}t}\;,
\label{approx}
\end{equation}
where we have introduced the Fourier
transform $\Delta({\bf k},{\bf R})$ of the order parameter 
$\Delta({\bf r},{\bf R})$. Since in the weak-coupling limit the pair
potential
is expected to be non-zero only close to $k_F$, one can replace the momentum by
an angle $\theta$.
Neglecting self-consistency, we therefore model the order parameter as
\begin{equation}
\Delta({\bf k},{\bf x},t)=
\left\{ \begin{array}{ll}
\Delta_1(\theta)\;, & x<-L/2 \\
0\;, & |x|<L/2 \\
\Delta_2(\theta)e^{i(\phi_0+2eVt/\hbar)}\;, & x>L/2. 
\end{array}
\right.
\label{ppot}
\end{equation}
Since
the overall phase is arbitrary, $\Delta_1$ is chosen real.

Before solving Eq.\ (\ref{tdBdG}), it is convenient to
introduce some notational simplifications. First we define the BCS coherence
factors
\begin{eqnarray}
&&\frac{v_i(E,\theta)}{u_i(E,\theta)}=\frac{E-\mbox{sgn}(E)\sqrt{E^2-\Delta_i(\theta)^2}}{\Delta_i(\theta)},\;
|E|>|\Delta_i(\theta)| \nonumber \\
&&\frac{v_i(E,\theta)}{u_i(E,\theta)}=\frac{E-i\sqrt{\Delta_i(\theta)^2-E^2}}
{\Delta_i(\theta)},\;
|E|<|\Delta_i(\theta)| 
\label{uv}
\end{eqnarray}
where $|u_i|^2+|v_i|^2=1$ and $i=1,2$ refers to superconductor 1 and 2. Using
this definition, one can express the Andreev reflection amplitude at
superconductor $i$ as 
$A_i(E_n,\theta)\equiv A_{i,n}=v_i(E_n,\theta)/u_i(E_n,\theta)$ and
$A_{i}(E_n,\bar{\theta})\equiv \bar{A}_{i,n}$, where $E_n=E+neV$.
The transmission amplitude for an electron-like quasiparticle at energy $E$ and
angle $\theta$ from superconductor $i$ to enter the electron branch in the
normal region is 
$J_{i}(E,\theta)\equiv J_i=
(u_{i}^2(E,\theta)-v_{i}^2(E,\theta))/u_{i}(E,\theta)$. 

Solving the scattering problem, we follow the approach by Averin and
Bardas.\cite{AveBar} First, we assume that an electron-like quasiparticle is
approaching the barrier from the left superconductor (labeled by
$\rightarrow$) at energy $E$ and 
angle $\theta$. Since we will calculate the current in the normal region to the
left of the barrier, see Fig.~\ref{fig1}, we need the wave function 
${\Psi}_{L}(E,\theta)\equiv {\Psi}_{L}$ that solves
Eq.\ (\ref{tdBdG}) in region "Left" (L), see Fig.\ \ref{fig1}:
\begin{equation}
{\Psi}_{L}^{\rightarrow} =
\sum_{n} 
\left\lgroup
\begin{array}{c} a_{n}^{\rightarrow}e^{i {\bf k}\cdot{\bf x}}
+ d_{n}^{\rightarrow} e^{i {\bf\bar{k}}\cdot{\bf x}} 
\\
b_{n}^{\rightarrow} e^{i {\bf k}\cdot{\bf x}} 
+c_{n}^{\rightarrow} e^{i {\bf \bar{k}}\cdot{\bf x}} 
\end{array}
\right\rgroup
e^{-i (\frac{E_n t}{\hbar}+\frac{n\phi_0}{2})} \; ,
\label{wfnL}
\end{equation}
where ${\bf k}=(k_x,k_y)=k(\cos\theta,\sin\theta)$, ${\bf \bar{k}}=(-k_x,k_y)
=k(\cos\bar{\theta},\sin\bar{\theta})$ and
$E_n=E+neV$ ($n$ is an even integer). 

Matching the wave functions of the
various regions at the two NS interfaces and the barrier, the following
(recursive) relations can be derived that determine the coefficients $a$,
$b$, $c$ and $d$. For $d$ we have
\begin{equation}
\alpha_n d_{n+2}+\beta_n d_{n}+\gamma_n
d_{n-2}=rJ_1\delta_{n,0},
\label{recd}
\end{equation}
where $\alpha_n$, $\beta_n$ and $\gamma_n$ are defined as
\begin{eqnarray}
&&\alpha_n=-|t|^2\frac{\bar{A}_{2,n+1}\bar{A}_{1,n+2}}
{1-A_{2,n+1}\bar{A}_{2,n+1}}
\nonumber \\
&&\beta_n=1-A_{1,n}\bar{A}_{1,n}+|t|^2\left(\frac{A_{1,n}\bar{A}_{1,n}}
{1-A_{2,n-1}\bar{A}_{2,n-1}}
\right.
\label{abc} \\
&&\left.+\frac{A_{2,n+1}\bar{A}_{2,n+1}}{1-A_{2,n+1}\bar{A}_{2,n+1}}\right)
\;, \ \ \ \ \ \gamma_n=-|t|^2\frac{A_{2,n-1}A_{1,n}}{1-A_{2,n-1}
\bar{A}_{2,n-1}}.  \nonumber
\end{eqnarray}
The coefficient $b$ is determined from
\begin{equation}
b_{n+2}=\frac{|t|^2A_{2,n+1}d_{n}+\bar{A}_{1,n+2}(|r|^2-A_{2,n+1}
\bar{A}_{2,n+1})d_{n+2}}{r(1-A_{2,n+1}\bar{A}_{2,n+1})}.
\label{b}
\end{equation}
The coefficients $a$ and $c$ are related to $b$ and $d$
through
\begin{equation}
a_{n}=A_{1,n}b_{n}+J_1\delta_{n,0},\ \ \ c_{n}=\bar{A}_{1,n}
d_{n}\;.
\label{ac}
\end{equation}
Using the wave function in Eq.\ (\ref{wfnL}) we can now calculate 
the current $I$ per $ab$-plane (repeating the calculation above for the
left-movers):  
\begin{eqnarray}
\frac{I}{\sigma_0}=&&\frac{1}{2T}\int_{-\pi/2}^{\pi/2}d\theta\cos\theta
\int_{-\infty}^{\infty}\frac{dE}{\Delta_0}\left[T^{\rightarrow}(E,\theta)-
T^{\leftarrow}(E,\theta)\right]\nonumber \\
&&+\frac{eV}{\Delta_0}\; ,\ \ \ \ \ \
\sigma_0=L_y\frac{2^{3/2}em^{1/2}E_{F}^{1/2}\Delta_0T}{h^2}\; ,
\label{curr}
\end{eqnarray}
where $T=\int d\theta |t|^2 \cos\theta/2$ and $L_y$ is the junction 
length in the $b$-direction. The quantities $T^{\tau}(E,\theta)$
are defined as ($\tau=\rightarrow,\leftarrow$) 
\begin{eqnarray}
T^{\tau}(E,\theta)=&&N_{\tau}(E,\theta)\left[f(E)
T^{\tau}_e(E,\theta)
-f(-E)T^{\tau}_h(E,\theta)\right],\nonumber \\ 
&&f(E)=1/[1+\exp(E/k_BT)],
\end{eqnarray}
where $N_{\tau}(E,\theta)$ is the bulk density of states evaluated in 
superconductor 1 (2) for $\tau=\rightarrow$ ($\leftarrow$),
and 
\begin{eqnarray}
&&T^{\tau}_e(E,\theta)=\sum_n\left[|a_{n}^{\tau}|^2-|d_{n}^{\tau}|^2\right]
,\nonumber \\ 
&&T^{\tau}_h(E,\theta)=\sum_n\left[|b_{n}^{\tau}|^2-|c_{n}^{\tau}|^2\right].
\end{eqnarray}

The Eqs.\ (\ref{recd})-(\ref{ac}) are the main technical results of this paper.
Together with Eq.\ (\ref{curr}) we will in the following use these equations to
numerically calculate the $I-V$ relation for some $d$-wave junctions. 
Putting $A=\bar{A}$
(valid for $s$-wave) and assuming $\Delta_1=\Delta_2$,
Eqs.\ (\ref{recd})-(\ref{ac}) will exactly transform to Eq.\ (5) of Ref.\
\onlinecite{AveBar}.

\begin{figure}
\centerline{\psfig{figure=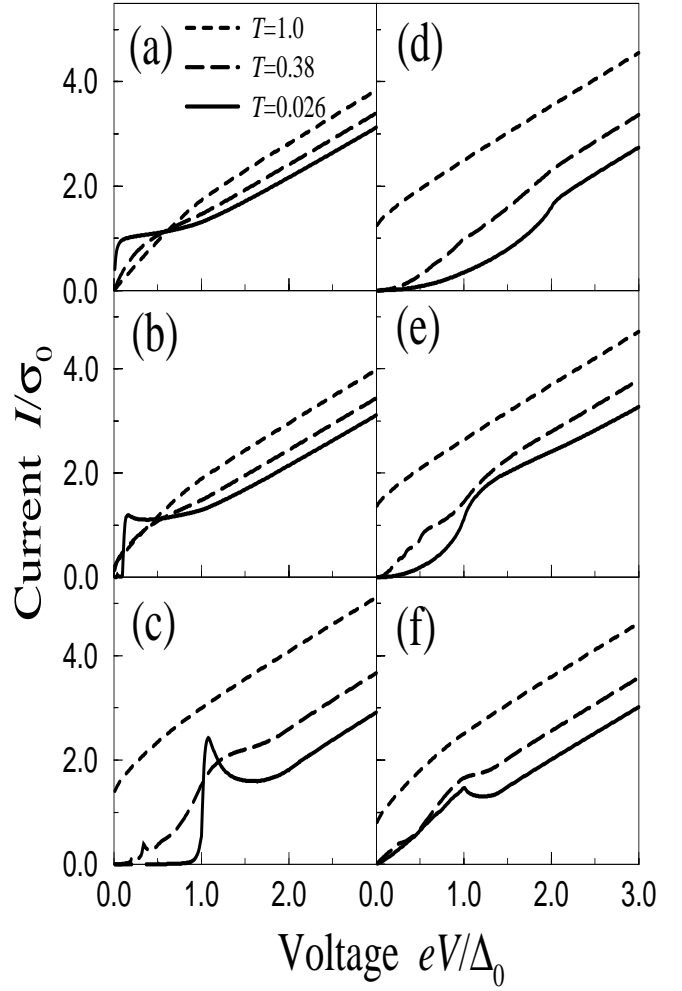,height=13cm,width=12cm}}
\vspace*{3mm}
\caption{
Varying the transmission $T$, we show $I-V$ curves for some arrangements of
$s$-wave and $d$-wave superconductors. In (a) we show the results for the 
N$/d_{xy}$ 
junction. In (b)-(f), the $I-V$ curves for junctions with superconducting
electrodes on both sides are shown: in (b) $s/d_{xy}$ with
$\Delta_s=0.1\Delta_0$; in (c) $s/d_{xy}$ with $\Delta_s=\Delta_0$; in (d)
$d_{x^2-y^2}/d_{x^2-y^2}$; in (e) $d_{xy}/d_{xy}$; in (f)
$d_{x^2-y^2}/d_{xy}$. Zero-temperature is assumed.
}
\label{fig2}
\end{figure}

In Fig.\ \ref{fig2} we present some calculations of $I-V$ curves involving 
$d$-wave superconductors. The MGS shows up in Eq.\ (\ref{abc}). The quantity
$1-A_{n}\bar{A_{n}}$ found in the denominators of Eq.\ (\ref{abc}) is zero when
$E_n=0$ for the case $\Delta(\theta)=-\Delta(\bar{\theta})$. This leads to
transmission resonances which we explore in the following. 

For the ballistic curves (transmission $T=1$, no MGS in this case) involving 
two superconductors there is
a non-vansishing zero-voltage current that disappears when the transmission is
not unity.\cite{HDB1,HDB2} 
In Fig.\ \ref{fig2}(a) we confirm that 
using the Eqs.\
(\ref{recd})-(\ref{curr}) reproduces previous results for the NS 
case.\cite{TanKas,XMT}
For low transmission, there is a huge increase of the 
current at small voltage leading to zero-bias conductance 
peaks.\cite{TanKas,XMT} 

Next, we study $s/d_{xy}$ junctions, see Figs. 2(b)-(c). As the $s$-wave gap 
$\Delta_s$ 
increases, the $s$-wave gap more and more "shadows" the zero-energy MGS on the 
$d$-wave side. Since the width of the MGS decreases with decreasing transmission
$T$ this effect is more pronounced for low transmission. 
Subharmonic gap structure (SGS) appears at voltages $\Delta_s/n$, $n>0$
integer, see Fig.~6(a) in Ref.\ \onlinecite{HDB1}. All other SGS predicted to 
occur in asymmetric 
junctions\cite{HDB1,HDB2} is partly washed out after angular averaging
since it involves also the gap of the $d$-wave superconductor (which depends on
$\theta$). The SGS at
voltages $\Delta_s/n$ depends on $n$ being even or odd. For odd $n$ the
scattering states originating from the left ($s$-wave) superconductor 
will hit the resonant MGS at the $d$-wave side, leading to negative
differential resistance. For even $n$ the scattering states responsible for
SGS do not hit the
MGS, and therefore this case resembles more what one finds in junctions with
$s$-wave superconductors on both sides. When the transmission decreases, all
SGS besides $n=1$ vanishes, in which case the negative differential conductance
is large.   

We also study the case when there are $d$-wave superconductors on both sides of
the junction. First, the case $d_{x^2-y^2}/d_{x^2-y^2}$ is shown in
Fig.~\ref{fig2}(d). In this case the MGS are absent since
$\Delta(\theta)=\Delta(\bar{\theta})$. We can therefore compare our
low-transmission curve [the case $T=0.026$ in Fig.~\ref{fig2}(d)] with the
tunneling calculation shown in Fig.~2 of Ref.\ \onlinecite{BOZ}, and we find 
that the curves
are the same. At small voltage $V$ there is a $V^2$ dependence of the current, 
and
one finds a threshold voltage at $eV=2\Delta_0$. For
intermediate transmission [the case $T=0.38$ in Fig.~\ref{fig2}(d)] SGS appears
at voltages $2\Delta_0/n$. The SGS is not as sharp as in the $s$-wave case
because of angular averaging. 

For the $d_{xy}/d_{xy}$ configuration, see Fig.~\ref{fig2}(e), there are MGS's 
on either side of the junction. The difference compared to the
$d_{x^2-y^2}/d_{x^2-y^2}$ case is that the presence of MGS lower the threshold
voltage to $eV\approx \Delta_0$. This behavior is similar to the 
case when a Breit-Wigner resonance is embedded in the normal 
region.\cite{JSBW}
The reason for the lower threshold voltage is that the MGS transmission 
resonances at the right (left) hand side are most pronounced for 
$\theta\approx \pi/4$. At this angle a quasiparticle coming from the left 
(right) 
experiences the gap $\Delta(\pi/4)=\Delta_0$, resulting in a threshold
voltage $eV\approx \Delta_0$ to make it possible for the quasiparticle from the
left (right) side to hit the MGS on the right (left) side. Angular averaging
smears out the threshold. For small voltage
$V$ there is a $V^2$ dependence of the current in the low-transmission limit. 
For
intermediate transmission SGS again shows up at $2\Delta_0/n$, somewhat washed 
out because of angular averaging. 
In the $s/d_{xy}$ case the negative
differential conductance comes from resonant conduction of the right-moving
scattering states through the MGS of the right-hand side. In the
$d_{xy}/d_{xy}$ case both left-movers and right-movers will pass through 
MGS (since
there are MGS on both sides), and the resonant contribution from both
left-movers and right-movers cancels. 
Therefore, there is in the $d_{xy}/d_{xy}$ case no
negative differential conductance. 

In Fig.~\ref{fig2}(f) we show the $I-V$ curves for the $d_{x^2-y^2}/d_{xy}$ 
case. This case is similar to the $s/d_{xy}$ case shown in 
Figs.~\ref{fig2}(b)-(c) since
$\Delta_1(\theta)=\Delta_1(\bar{\theta})$. One important difference is that the
angular averaging now results in 
non-zero current for small voltages.
Still there is negative differential conductance close to the voltage 
$eV\approx \Delta_0$ in the low-transmission limit. One could understand the
$I-V$ curve in Fig.~\ref{fig2}(f) as an averaging of the curves in
Figs.~\ref{fig2}(a)-(c). For small transmission the current is proportional to
the voltage $V$ in the small-voltage limit.

In summary, we have calculated $I-V$ curves for 
planar $d$-wave junctions. 
In some arrangements we find considerable negative differential conductance and
lower threshold voltage for non-zero current due to resonant conduction through
MGS. These features have not been discussed before in the literature and
they should be experimentally
observable if superconducting gaps with $d$-wave symmetry exist
in nature.

I thank M. Samanta for many discussions on $d$-wave 
superconductivity and for a lot of inspirational energy to carry this project 
to an end. Also, I have benefitted from discussions with 
P. F. Bagwell, Z. Ivanov, G. Johansson, V. S. Shumeiko, G. Wendin and 
S. \"Ostlund.


\begin{references}
\bibitem{Harl}D. J. Van Harlingen, Rev. Mod. Phys. {\bf 67}, 515 (1995) and references therein.
\bibitem{Woll}D. A. Wollman {\it et al.}, Phys. Rev. Lett. {\bf 71}, 2134 
(1993).
\bibitem{Tsue}C. C. Tsuei {\it et al.}, Phys. Rev. Lett. {\bf 73}, 593 (1994).
\bibitem{ChauLin}P. Chaudhari and S.-Y. Lin, Phys. Rev. Lett. {\bf 72}, 1084 
(1994).
\bibitem{Sun}A. G. Sun {\it et al.}, Phys. Rev. Lett. {\bf 72}, 2267 (1994).
\bibitem{Hu}C.-R. Hu, Phys. Rev. Lett. {\bf 72}, 1526 (1994).
\bibitem{TanKas}Y. Tanaka and S. Kashiwaya, Phys. Rev. Lett. {\bf 74}, 3451 
(1995).
\bibitem{XMT}J. H. Xu, J. H. Miller, and C. S. Ting, Phys. Rev. B. 
{\bf 53}, 3604 (1996).
\bibitem{BSW}E. N. Bratus, V. S. Shumeiko, and G. Wendin, Phys. Rev. 
Lett. {\bf 74}, 2110 (1995).
\bibitem{AveBar}D. Averin and A. Bardas, Phys. Rev. Lett. {\bf 75}, 
1831 (1995).
\bibitem{WenOst}F. Wenger and S. \"Ostlund, Phys. Rev. B {\bf 47}, 5977 
(1993).
\bibitem{DevFul}T. P. Devereaux and P. Fulde, Phys. Rev. B {\bf 47}, 14 638
(1993).
\bibitem{BOZ}C. Bruder, A. van Otterlo, and G. T. Zimanyi, Phys. Rev. 
B {\bf 51}, 12 904 (1995).
\bibitem{BGZ}Yu. S. Barash, A. V. Galaktionov, and A. D. Zaikin, Phys. Rev. 
B {\bf 52}, 665 (1995).
\bibitem{HDB1}M. Hurd, S. Datta, and P. F. Bagwell, Phys. Rev. B {\bf 54},
6557 (1996).
\bibitem{HDB2}M. Hurd, S. Datta, and P. F. Bagwell (unpublished).
\bibitem{Brud}C. Bruder, Phys. Rev. B. {\bf 41}, 4017 (1990).
\bibitem{JSBW}G. Johansson {\it et al.}, in
Proceedings of the 1997 International Symposium on Intrinsic Josephson Effect
and THz Plasma Oscillations in High $T_c$ Superconductors, Sendai, Japan, 
23-25 February 1997 (to be published).
\end{references}
\end{document}